\documentclass[journal,comsoc]{IEEEtran}
\usepackage{units}

\usepackage{graphicx}
\usepackage{amsmath,amsfonts,amsthm,bm} 

\usepackage{amssymb}
\usepackage{amsxtra}
\usepackage[cmintegrals]{newtxmath}
\usepackage{amstext}
\usepackage{amssymb}
\usepackage{latexsym}
\usepackage{afterpage}
\usepackage{verbatim}
\usepackage{psfrag}
\usepackage{color}
\usepackage[table]{xcolor}
\usepackage{cite}
\usepackage{times}
\usepackage{wrapfig}
\usepackage{pgfplots}
\usepackage{multirow}
\usepackage{tikz}
\usepackage{setspace}
\usepackage{epsfig}
\usepackage{epstopdf}
\usepackage{footmisc}
\usepackage{fmtcount}
\usepackage{stackrel}
\usepackage{float}
\usepackage{breqn}
\usepackage{enumerate}
\usepackage{subcaption}
\usepackage{hyperref}
\usepackage{hypcap}
\usepackage[ruled,vlined,linesnumbered]{algorithm2e}
\usepackage[nottoc]{tocbibind}
\usepackage[normalem]{ulem}
\usepackage{authblk}
\usepackage{tabularx}
\usepackage{bbm}
\usepackage{array}
\usepackage{mathtools}
\usepackage{listings}

\usepackage{mathtools}

\newcommand{\SPP}{\mathrm{SPP}}
\begin{document}

\title{Ultra-Massive MIMO Systems at Terahertz Bands: Prospects and Challenges}

\author{Alice~Faisal,~\IEEEmembership{Student Member,~IEEE,}
        Hadi~Sarieddeen,~\IEEEmembership{Member,~IEEE,}
        Hayssam~Dahrouj,~\IEEEmembership{Senior Member,~IEEE,}
        Tareq~Y.~Al-Naffouri,~\IEEEmembership{Senior Member,~IEEE}
        and~Mohamed-Slim~Alouini,~\IEEEmembership{Fellow,~IEEE}
\thanks{This paper is accepted at the IEEE Vehicular Technology Magazine (IEEE Future Networks Initiative Series on 6G Technologies and Applications).

This research is supported by the King Abdullah University of Science and Technology (KAUST) Office of Sponsored Research, and in part by the Center of Excellence for NEOM Research at KAUST.

An early version of this work was presented at the 43rd Wireless World Research Forum (WWRF43) meeting in London, UK, Oct. 2019  \cite{faisal2019WWRF}. 

A. Faisal is with the Department of Electrical Engineering, Effat University, Jeddah 22332, Saudi Arabia (e-mail:alfaisal@effat.edu.sa; hayssam.dahrouj@gmail.com).

H. Sarieddeen, T. Y. Al-Naffouri, and M.-S. Alouini are with the Division of Computer, Electrical and Mathematical Sciences and Engineering, King Abdullah University of Science and Technology, Thuwal 23955-6900, Saudi Arabia (e-mail: hadi.sarieddeen@kaust.edu.sa; tareq.alnaffouri@kaust.edu.sa; slim.alouini@kaust.edu.sa).

H. Dahrouj is with the Center of Excellence for NEOM Research, King Abdullah University of Science and Technology,  Thuwal 23955-6900, Saudi Arabia (e-mail: hayssam.dahrouj@gmail.com).

}
}

\maketitle

\begin{abstract}
Terahertz (THz)-band communications are currently being celebrated as a key technology that could fulfill the increasing demands for wireless data traffic in the upcoming sixth-generation (6G) of wireless communications. Many challenges, such as high propagation losses and power limitations, which result in short communication distances, have yet to be addressed for this technology to be realized. Ultra-massive multiple-input, multiple-output (UM-MIMO) antenna systems have emerged as practical means for combatting this distance problem, thereby increasing system capacity. Towards that end, graphene-based nano-antennas have recently been proposed, as they can be individually tuned and collectively controlled in compact UM-MIMO array-of-sub-arrays architectures. In this paper, we present a holistic overview of THz UM-MIMO systems. We assess recent advancements in transceiver design and channel modeling, and discuss the major challenges and shortcomings of such designs by deriving the relationship between communication range, array dimensions, and system performance. We further highlight several research advances that could enhance resource allocation at the THz band, including waveform designs, multi-carrier configurations, and spatial modulations. Based on this discussion, we highlight prospective use cases that can bring THz UM-MIMO into reality in the context of sensing, data centers, cell-free systems, and mid-range wireless communications.

\end{abstract}

\IEEEpeerreviewmaketitle

\section{Introduction}
\label{sec:Intro}

\IEEEPARstart{T}{erahertz} (THz)-band communications promise to exploit the large bandwidths at THz frequencies in order to fulfill the high data rate demands of the future sixth-generation (6G) of wireless mobile communications and beyond \cite{8732419Rappaport,sarieddeen2019generation}. In recent years, millimeter-wave (mmWave) systems have been extensively explored over the frequency range of 30 to 300 gigahertz (GHz). Since the highest attainable bandwidth at such frequencies is a few tens of GHz, a physical-layer efficiency of hundreds of bits per second per Hertz (bits/sec/Hz) is required to achieve a terabit (Tb)-per-second data rate. Since the available bandwidth between 0.3 THz and 10 THz (i.e. at the THz range) can reach hundreds of GHz, a target Tb/sec data rate can be achieved with minimal physical-layer efficiency-enhancement techniques \cite{akyildiz2016realizing}.

\begin{figure}[t]
\centering
\includegraphics[width=3.5in]{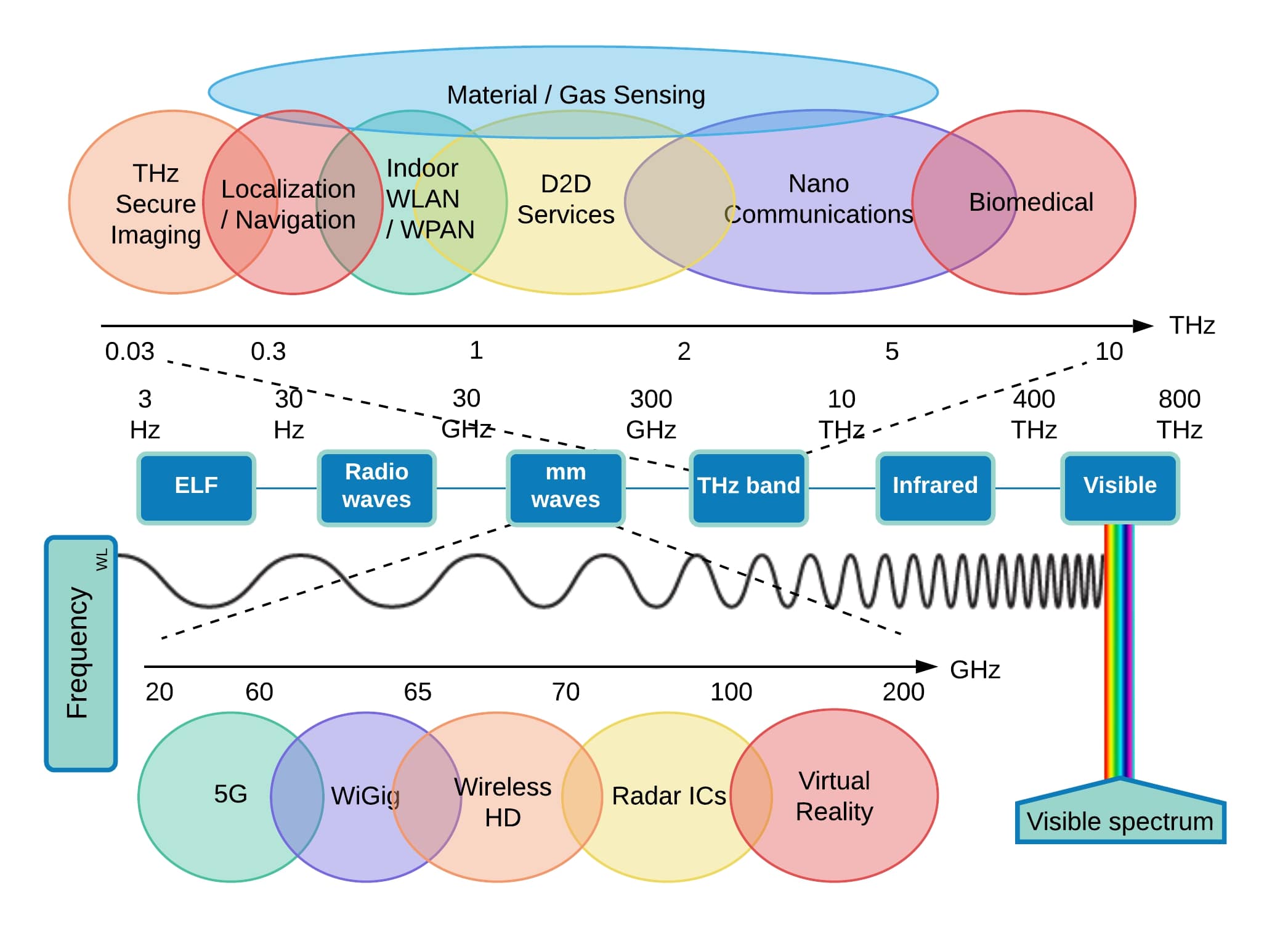}
\caption{An illustration of the spectrum decomposition and the corresponding high-frequency applications.}
\label{f:spectrum}
\end{figure}

Over the past few years, affordable technologies have enabled widespread use of mmWave systems, which are now the essential enablers of the fifth-generation (5G) mobile communication networks. For example, mmWave-enabled IEEE 802.11ad (WiFi) networks (WiGig), high-definition (HD) video applications, and single-chip, fine-resolution, radar-integrated circuits have emerged. 
Recently, major advancements in transceiver design are closing the so-called THz gap, paving the way for applications at the THz band ranging from indoor wireless communications to vehicular and drone-to-drone communications, device-to-device (D2D) communications, and nano-communications. Furthermore, THz signals have the potential to be used in many non-communication-based applications, such as in the spectroscopy of small bio-molecules and in the quality control of pharmaceutical products. Spectrum decomposition and its corresponding applications are illustrated in Fig. \ref{f:spectrum}. 

Despite the promising features of THz communications systems, their high-frequency operation properties impose several implementation hurdles, both at the signal-generation and the signal-detection levels. To address such implementation challenges, we propose a variety of integrated electronic and photonic THz solutions. These solutions do not necessarily result in perfect devices, but rather in programmable devices that are efficient and that satisfy emerging system-level properties (see \cite{sengupta2018terahertz} and references therein). 
At high-frequency ranges, however, electronic
generators and detectors perform poorly at room temperature (plasma waves excited by high-electron mobility transistors, for example, tend to be  unstable). On the other hand, photonic solutions using photomixers and difference-frequency generation, for example, suffer from a limited output power.

Recently, plasmonic solutions --in particular, graphene-based solutions \cite{ju2011graphene}-- have emerged as strong candidates that enable communications at the THz band. The unique electrical properties of graphene --such as high electron mobility, electrical tunability, and configurability-- allow supporting high-frequency signals. Plasmonic-based antennas enable the propagation of surface plasmon polariton (SPP) waves.
In fact, SPP waves propagate at speeds that are much lower than those of regular EM waves and, hence, possess a characteristic wavelength (denoted by $\lambda_{\SPP}$) that is much smaller than the EM wavelength (denoted by $\lambda$). Therefore, compact array designs, which integrate a massive number of antennas in a tiny footprint, can be deployed \cite{akyildiz2016realizing}.  The confinement factor, which defines the number of AEs that can be embedded in a fixed footprint, is denoted by ($\gamma$ = $\lambda/\lambda_{\SPP}$). Generally, the confinement factor in graphene is between 10 and 100, which makes THz-operating plasmonic antennas almost two orders of magnitude smaller than metallic antennas.

Additional crucial challenges still need to be addressed from a coverage perspective, particularly those related to high propagation losses and power limitations. To overcome such limitations, we distinguish between reflect arrays and ultra-massive multiple-input, multiple-output (UM-MIMO) antenna systems as alternative means to extend the coverage range \cite{akyildiz2016realizing}. In order to mimic massive MIMO systems, reflect arrays can further be used on the transmitter side to generate a directional THz beam excited by a THz source in the proximity. UM-MIMO offers the valuable advantages of increasing the communication range and improving the achievable data rate through spatial multiplexing (SMX) and beamforming, respectively.

The main aim of this paper is to establish a clear link between the transceiver design, the channel characteristics, and the prospective use cases of THz UM-MIMO systems. To our knowledge, the literature lacks a holistic work of this kind. Towards this end, then, in Sec. \ref{sec:circuit}, we introduce the array-of-sub-arrays (AoSA) configuration and provide a comparison between the antenna compactness of mmWave and THz-antenna arrays as well as their achievable SMX gains. In Sec. \ref{sec:chmodel}, we detail various channel-modeling approaches.
We further define open challenges and potential research advances in Sec. \ref{sec:advantages}. Finally, based on previously-discussed constraints, we recommend specific THz UM-MIMO use cases in Sec. \ref{sec:prop_sol} and conclude in Sec. \ref{sec:conclusion}.

\section{Array of Sub-Arrays Design}
\label{sec:circuit}

 \begin{figure*}[t]
        \centering
        \begin{subfigure}{0.492\textwidth}
            \centering
            \includegraphics[width=\textwidth]{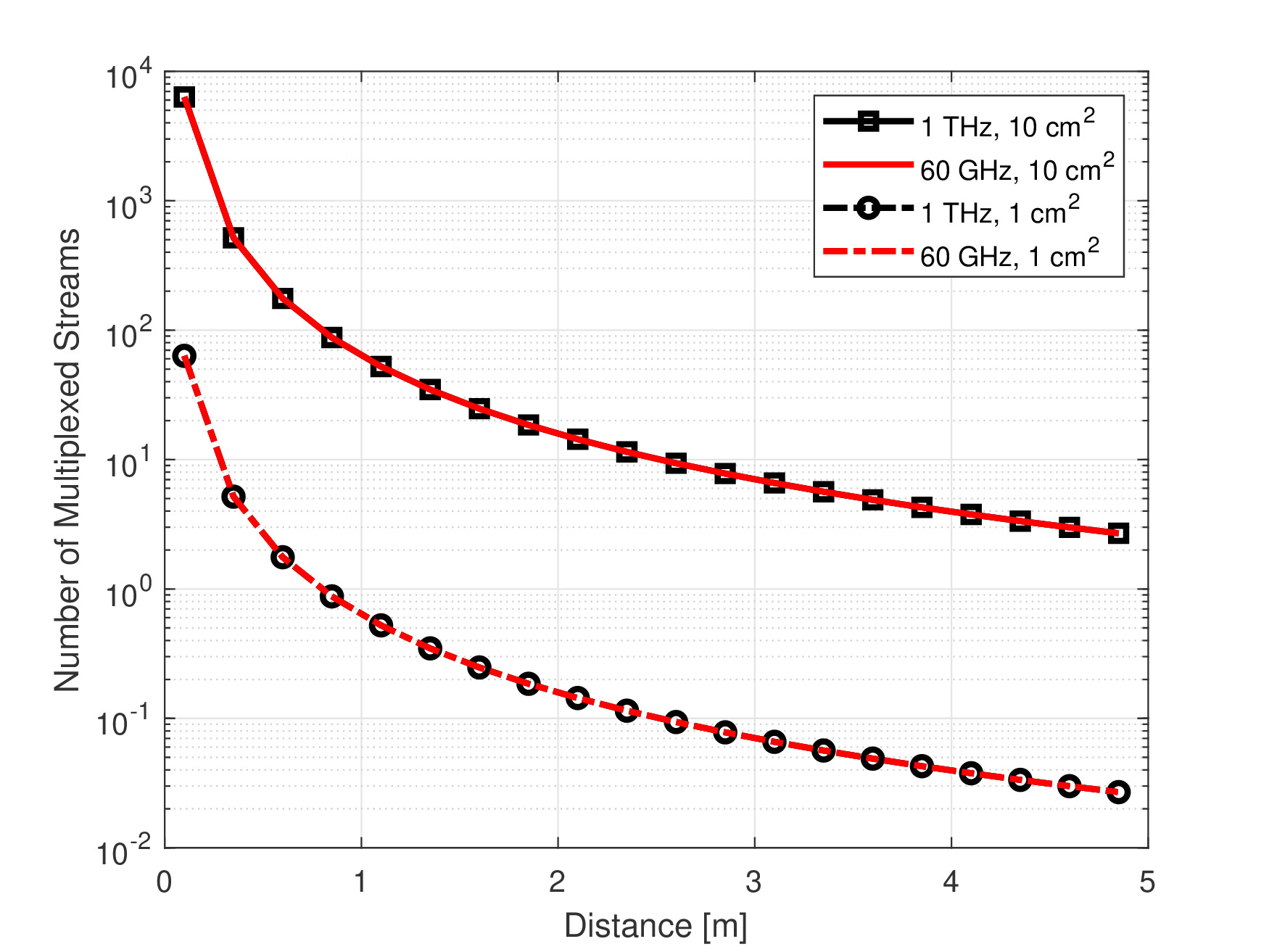}
            \caption{Metallic antenna arrays.}  
        \end{subfigure}
        \hfill
        \begin{subfigure}{0.5\textwidth}  
            \centering 
             \includegraphics[width=1\linewidth]{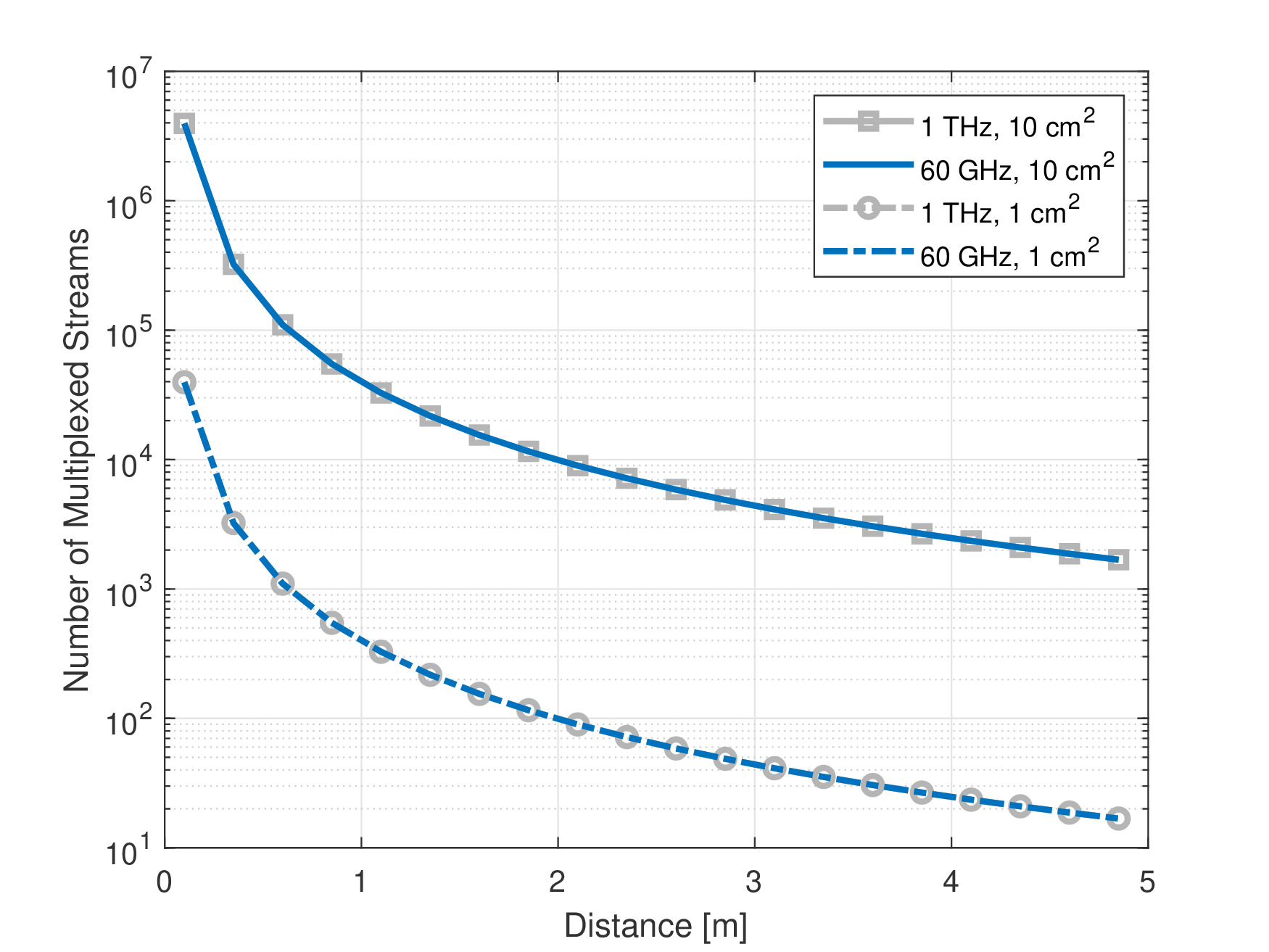}
     \caption{Graphene antenna arrays.} 
        \end{subfigure}
        \caption{Achievable multiplexing gains versus communication range in mmWave- and THz-antenna arrays.}
        \label{f:mmWave_vs_THz}
    \end{figure*}

Massive plasmonic antenna configurations are constructed as large arrays of antenna elements (AEs). Since inter-AE separations are typically on the order of $\lambda$, operating at high frequencies naturally results in dense packaging. For instance, while mmWave AoSAs require footprints of few square centimeters for a small number of antennas, at THz, a large number of antennas can be embedded in a few square millimeters \cite{akyildiz2016realizing}. This densification is increased with plasmonic antennas. This compactness in design, however, comes at the cost of limiting the beamforming and multiplexing gains of UM-MIMO because of inadequate spatial sampling and the increasing complexity of antenna-array control \cite{Sarieddeen8765243}. As a solution, in AoSA architectures, large antenna arrays can be divided into multiple sub-arrays (SAs) of smaller size. Deploying multiple AEs in an SA improves the beamforming gain and decreases the required transmission power for each element. Such AoSA configurations can guarantee a large number of directed, independent paths, each of which can be used to carry independent information. Hence, each SA offers the array gain, and the collaboration between SAs provides the SMX gain. The number of SAs can be expressed as 
\begin{equation}
    N = A \frac{\gamma \alpha^2 T_p }{\lambda^2 R_p}
    \end{equation}
where A, $T_p$, $R_p$, and $\alpha$ denote the array footprint, transmitted power, received power, and path loss, respectively. This is derived simply by dividing the total number of AEs in an array ($A (\gamma/\lambda)^2$) by the number of AEs in each SA ($R_p/T_p \alpha^2$).

A comparison between the maximum number of multiplexed streams in mmWave and THz antenna arrays is illustrated in Fig. \ref{f:mmWave_vs_THz}, which plots the number of multiplexed streams as a function of the communication range. The simulation assumes that a single dominant ray exists between the transmitting and receiving SAs (the line-of-sight (LoS) path). It further considers a single-carrier MIMO system and assumes a transmission power of $\unit[10]{mW}$ and a noise power of $\unit[-80]{dBm}$ on the receiver side. In order to meet a specific communication distance with the required receive sensitivity, a specific number of AEs per SA is required to generate a sufficiently large beamforming gain. Under the same transmit power capabilities, Fig. \ref{f:mmWave_vs_THz} shows that the multiplexed streams in same-size mmWave and THz arrays are identical. Although the path loss increases by the square of the wavelength, the number of antennas that can be fit in the same footprint also increases by the square of the wavelength. Hence, the obtained beamforming gains would compensate for the increased path loss. Furthermore, identical number of streams indicates an identical number of required radio frequency (RF) chains in a fixed hybrid AoSA architecture. The true achievable SMX gains, however, depend on the exact channel conditions, where spherical wave propagation at THz frequencies could improve such gains. While the above result justifies the intuition that mmWave and THz UM-MIMO systems are fundamentally similar from a high-level design perspective, much higher data rates are provided through THz chains. Such gains, however, are predicated on the peculiarities of THz-transceiver design and its corresponding impairments (e.g., RF non-idealities, phase noise, carrier-frequency offsets, power-amplifier saturation, etc.), which are expected to eventually define THz band UM-MIMO systems.

Since the beamforming strategy dictates the energy and spectral efficiencies in an AoSA architecture, hybrid beamforming, in which operations get divided between the analog and digital domains, is typically sought in order to reduce hardware costs and power consumption. Hybrid AoSA architectures at THz are detailed in \cite{7786122}, which illustrates a two-step analog beamforming and user grouping mechanism based on users' angle of departure. The beamforming angle is selected such that the overall received signal power is maximized for each user group over all subcarriers. Afterwards, digital beamforming is performed in baseband on each subcarrier. 

\begin{figure}[t]
\centering
\includegraphics[width=3.4in]{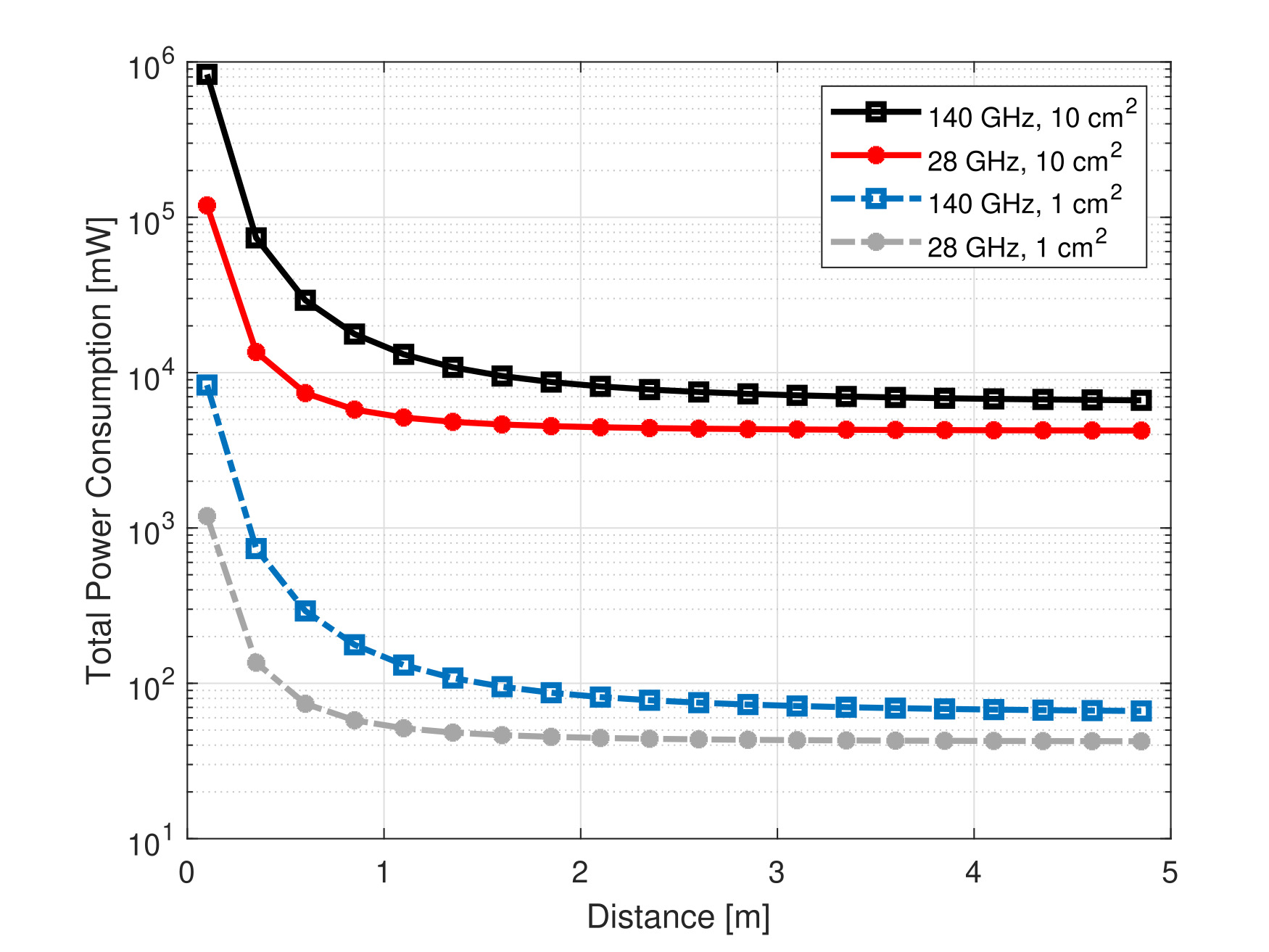}
\caption{Power consumption in AoSA architectures for different footprints and frequencies of operation.}
\label{f:power}
\end{figure}

 Power consumption is also a critical metric that needs to be considered for the practical deployment of mmWave and THz systems. Figure \ref{f:power} shows the power consumption of AoSA architectures at mmWave ($\unit[28]{GHz}$) and sub-THz ($\unit[140]{GHz}$) frequencies. Metrics on the power consumption of true THz components are still lacking at these frequencies. Our analysis assumes a single-carrier system and estimates the power dissipated by three major power-consuming components --namely, the low-noise amplifier (LNA) per AE, the analog-to-digital converter (ADC) per SA, and the mixer per SA--. State-of-the-art device characteristics are used to estimate the power consumption of such emerging systems, as reported in \cite{power}. The power consumed by the LNA is defined as
\begin{equation}
P_{\text{LNA}} = \frac{G}{(NF-1)\text{FoM}},
\end{equation}
where $G$, FoM, and $NF$ denote the gain (which should also account for the insertion loss of the phase shifter per antenna), figure of merit in $\unit[]{mW^{-1}}$, and noise figure, respectively. Each LNA consumes $\unit[11.36]{mW}$ at 28 GHz and $\unit[49.75]{mW}$ at 140 GHz. The power consumption drawn by each ADC is given by
\begin{equation}
    P_{\text{ADC}} =  f_{\text{samp}}2^n \text{FoM},
\end{equation}
where FoM, $f_{\text{samp}}$, and $n$ denote the figure of merit for the data conversion, sampling rate, and number of bits, respectively. We note an exponential increase in power consumption with the number of bits and linear increase with bandwidth. Each ADC consumes $\unit[8.17]{mW}$ at $\unit[28]{GHz}$ and $\unit[32.71]{mW}$ at $\unit[140]{GHz}$, respectively. Moreover, based on \cite{power}, the power drawn by the mixer is $\unit[10]{dBm}$ for $\unit[28]{GHz}$ systems and $\unit[19.9]{dBm}$ for $\unit[140]{GHz}$ systems. Figure. \ref{f:power} illustrates the total AoSA power consumption for mmWave and sub-THz frequencies in $\unit[1]{cm^2}$ and $\unit[10]{cm^2}$ footprints. In both cases, the power consumption of a $\unit[140]{GHz}$ receiver is much higher than that of a $\unit[28]{GHz}$ receiver, which indicates that considerable efforts are required to improve the power efficiency in high-frequency devices.

\section{Channel Modeling and Characteristics}
\label{sec:chmodel}

The previous section discusses how UM-MIMO AoSAs can combat propagation losses and maximize the achievable gains in the THz band. The exact performance of THz UM-MIMO systems, however, is dictated by the exact channel conditions and the corresponding accuracy in channel state information. Accurate channel models are thus a prerequisite for efficient use of the THz band. Such models should take into account the impact of both spreading and molecular absorption losses. Furthermore, LoS, NLoS, reflected, and scattered paths should be considered, and static and time-variant environments should be treated separately. 
In what follows, we review several channel-modeling approaches and we detail the peculiar characteristics of the THz channel.

\subsection{THz Channel Modeling}

THz channel-modeling approaches are deterministic, statistical, or hybrid \cite{Han8387210}. Deterministic channel modeling depends on site geometry and is often accomplished via ray-tracing (RT) techniques that are capable of handling site-specific structures. Applying RT to every channel path, however, increases the system complexity. As a solution, point-to-point RT can first be used to capture the losses between virtual points at the transmitter and at the receiver, and the resultant model can then be mapped to other AEs, reducing computational complexity. Statistical modeling is either matrix-based or reference-antenna-based. In a matrix-based model, each independent sub-channel is represented by a complex Gaussian variable. Reference-antenna-based models assume single-input, single-output statistical propagation for two reference antennas with array-steering vectors at the transmitter and the receiver. Finally, hybrid channel modeling combines the advantages of both deterministic and statistical approaches, in which dominant paths can be individually captured by the deterministic method, while other paths can be statistically generated. The latter approach captures the spatiotemporal properties while allowing smooth time evolution and avoiding channel discontinuity. THz channel models should further account for challenges caused by the relatively large dimensions of massive antenna arrays. In particular, non-stationarities arise when multiple regions of the array visualize different propagation paths. Furthermore, with large array dimensions, it is difficult to fulfill the plane-wave assumption, especially with non-plasmonic arrays. In this case, spherical wave-front channel-modeling approaches should be considered instead.\\

\begin{figure}[t]
\centering
\includegraphics[width=3.5in]{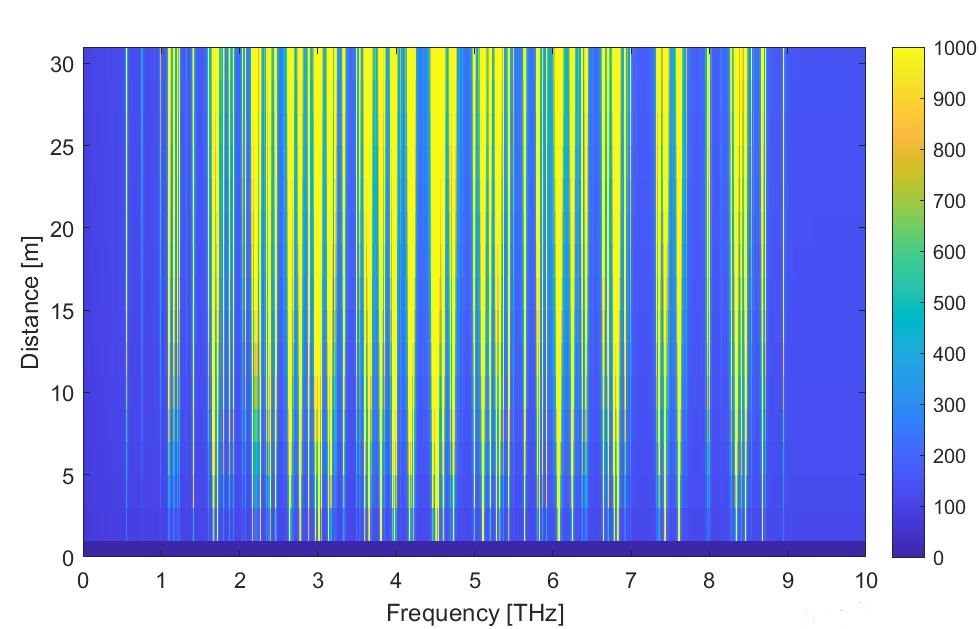}
\caption{Path loss from water-vapor molecules as a function of the communication range and frequency of operation.}
\label{f:channel_1}
\end{figure}

\begin{table}[t]
\small
\caption{Simulation Parameters and their Typical Values ($T$: system temperature, $p$: system pressure, $q^{i,g}$: mixing ratio of (isotope,gas)} $(i,g)$, $f_{c0}^{i,g}$: resonant frequency of $(i,g)$ at reference pressure, $\gamma$: temperature broadening coefficient, $\delta^{i,g}$: linear pressure shift of $(i,g)$, $S^{i,g}$: line intensity, $\alpha^{air}_0$: broadening coefficient of air, $\alpha^{i,g}_0$: broadening coefficient of $(i,g)$)\label{table:para}
\centering
 \begin{tabular}{| c || c |} 
 \hline
 Parameter & Value  \\ %[0.5ex] 
 \hline\hline
$T$ & $\unit[396]{K}$ \\ \hline
$p$ & $\unit[1]{atm}$ \\ \hline
$q^{i,g}$ & $\unit[0.05]{}$ \\ \hline
$f_{c0}^{i,g}$ & $\unit[(0 \sim 276.45)]{Hz}$ \\ \hline
$\gamma$ & $(-0.16 \sim 0.83)$ \\ \hline
$\delta^{i,g}$ & $\unit[(-0.0409 \sim 0.0251)]{Hz}$ \\ \hline
$S^{i,g}$  & $\unit[(9.98^{-36} \sim 2.66^{-18})]{Hz\  m^2/molecule}$ \\ \hline
$\alpha^{air}_0$ & $\unit[(0.0023 \sim 0.1117)]{Hz}$ \\ \hline
$\alpha^{i,g}_0$ & $\unit[(0.052 \sim 0.916)]{Hz}$ \\ \hline
 \hline
 \end{tabular}
\end{table}

\subsection{THz Channel Conditions}
\label{thz_cond}

At THz frequencies, the channel response is dominated by molecular absorption losses. The LoS path loss from water vapor molecules is between $\unit[0.1]{THz}$ and $\unit[10]{THz}$ over a distance range of $\unit[30]{m}$, illustrated in Fig. \ref{f:channel_1}. Note that the plot is dominated by spikes (in yellow) that originate from excited-molecule vibrations at specific resonant frequencies within the THz band. With certain spikes only appearing at specific distances, the available spectrum is divided into smaller distance-dependent windows. 
In fact, molecular absorption results in a more severe spectrum shrinkage at larger communication distances due to broadened absorption spikes and higher cumulative absorption losses.
All the parameters for absorption-loss computations can be obtained from the high-resolution transmission molecular-absorption database (some of which are summarized in Table \ref{table:para}); detailed mathematical expressions for computing the absorption cross-section can be found in \cite{Jornet5995306}. Note that molecular absorptions are followed by coherent re-radiations that can be lumped in with channel-induced absorption noise factors.

Because of high reflection losses, the THz channel mainly consists of a single LoS path and a few NLoS paths; the effect of scattered and refracted rays is negligible. Furthermore, the channel tends to be sparse with beamforming and ill-conditioned with SMX. Nevertheless,
achieving good multiplexing gains in high-frequency point-to-point LoS environments is feasible when antenna spacings are much larger than the operating wavelength. 
At THz frequencies, three regions of operation can be observed, illustrating the relationship between system performance, communication range $D$, and array dimensions:
\begin{itemize}

	\item In \emph{Region 1}, the distance $\Delta$ between SAs is relatively large, and the channel is always well-conditioned; this is a typical large-intelligent-surface (LIS) setting. 
	
	\item In \emph{Region 2}, $D$ is very large compared to $\Delta$ and the channel is highly correlated. By finely tuning $\Delta$, however, reference \cite{Sarieddeen8765243} shows that operations in \emph{Region 2} can be carried on eigenchannels to achieve efficient spatial multiplexing and spatial modulation (SM).
	
	\item In \emph{Region 3}, $D$ is much larger than the so-called \emph{Rayleigh distance}, where \emph{Region 2} optimizations no longer hold because of the limited physical sizes of antenna arrays. LoS communications beyond the \emph{Rayleigh distance} are studied in the context of mmWave communications in \cite{6800118Wang}. 	
\end{itemize}
 Note that, under spherical wave assumptions, the boundaries of such regions need to be rectified. Furthermore, \cite{8732419Rappaport} proposes spatial oversampling, which consists of significantly reducing the AE separations $\delta$
such that the spatiotemporal frequency-domain region of support for plane waves is decreased. Spatial noise shaping is one possible advantage of such oversampling, which helps in reducing noise figures, increasing linearity, and relaxing the high-resolution requirements of data converters. 

\section{Recent Research Advances}
\label{sec:advantages}

Due to fundamental differences in signal and channel characteristics, classical signal-processing problems have to be readdressed at the THz band. Such problems include, but are not limited to, accurate beamforming and beamsteering criteria, optimal precoding and combining methods, low-cost channel-estimation paradigms, and near-optimal data detection. 
We hereby highlight some relevant research advances.

\subsection{Modulation}
\label{sec:pulse-based-modulation}

The limitations of nano-scale transceivers bound their ability to generate continuous carrier-based modulations. In fact, with graphene at room temperature, only very short pulses with a corresponding power of few milliwatts can be generated in the higher THz range. This does not support long-distance communications. To address this limitation, pulse-based, asymmetric, on-off-keying modulation spread-in-time is investigated in \cite{Jornet6804405}. This modulation scheme transmits short one-hundred-femtosecond-long pulses for a logic pulse one, and it can support a very large number of nano-devices with up to a Tb/sec data rate. At least for the time being, most of the algorithms that are tailored for regular MIMO systems should be modified to account for pulse-based modulations.

\subsection{Waveform Design and Beamforming}
\label{sec:distance_aware}

In order to make the best use of spectral windows, THz-specific, optimized, multi-carrier waveform designs are required besides orthogonal frequency-division multiplexing. Since the channel is assumed flat, such designs would typically be single-carrier, with the possibility of incorporating carrier aggregation schemes. Nevertheless, due to some multipath components, frequency selectivity can still occur in THz indoor environments, and THz receivers can themselves be frequency selective. An efficient use of resources can be maintained through optimization frameworks that jointly control transmission power, sub-window allocation, and modulation formats. Similarly, efficient beamforming schemes are required so as to overcome the high path loss and account for the distance-dependent and frequency-dependent THz-channel characteristics. 
Furthermore, the THz channel's path components can split into different spatial directions for different sub-carrier frequencies or different frequencies within a subcarrier. This beam-squint effect results in a considerable array-gain loss. Since such losses are more significant in analog architectures, novel hybrid precoding architectures should be developed to mitigate the resultant beam-squint. Novel signal-processing techniques, such as sparse factorizations of delay Vandermonde matrices \cite{8974228Perera}, for example, can enable beam-squint-free communications.

\subsection{Multi-Carrier Antenna Configurations}
\label{sec:multicarrier}

In a plasmonic AoSA architecture, nano-antenna spacings can be significantly reduced, to the order of $\lambda_{\SPP}$, while still avoiding mutual coupling effects. Placing AEs very close to each other, however, limits the achievable gains by reducing the corresponding spatial-sampling capabilities. In fact, the maximum distance separation $\delta$ between two AEs should be on the order of half the operating wavelength $\lambda/2$ to avoid grating-lobe effects in beamforming. Towards that end, an interleaved antenna map \cite{Sarieddeen8765243}, in which neighboring AEs operate at different absorption-transmission windows, can be used. 

\begin{figure}[t]
\centering
\includegraphics[width=3.28in]{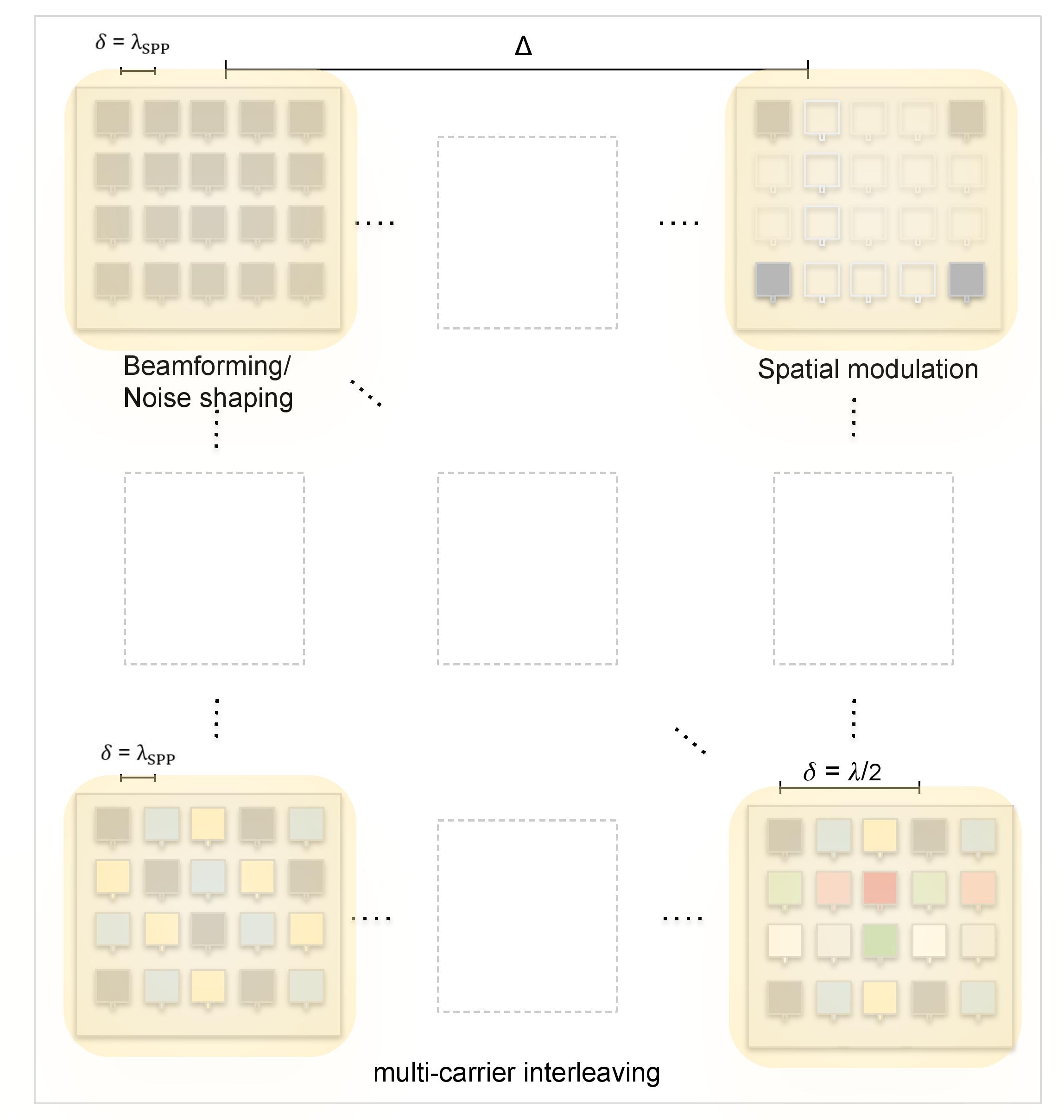}
\caption{An illustration of interleaved AoSA structures at the level of AEs and SAs.}
\label{f:AoSAs}
\end{figure}

Similarly, on the SA level, optimal separations $\Delta$ between same-frequency SAs can be maintained to achieve significant multiplexing gains (as argued in Sec. \ref{thz_cond}). 
Figure \ref{f:AoSAs} illustrates different interleaving schemes, in which AEs having the same color operate at the same frequency. The separation between two AEs tuned to the same frequency is $\delta\!=\!\lambda/2$, and that between two AEs tuned to different frequencies is $\delta\!=\!\lambda_{\SPP}$.  The AoSA structure further promotes different use cases, including noise shaping and spatial modulation, for densely-packed AEs. Note that if each AE is individually powered, larger array gains can be achieved. 
Furthermore, in wideband scenarios, the time of arrival for different paths at different SAs varies, resulting in inter-symbol interference. To mitigate such spatial-frequency wideband effects \cite{8732419Rappaport}, the channel models have to maintain spatial consistency.

\subsection{Spatial and Index Modulation}
\label{sec:SM}

SM can be thought of as a spectrum-efficient and power-efficient solution for THz UM-MIMO. Instead of antenna frequency maps, like those shown in Fig. \ref{f:AoSAs} for multi-carrier designs, we can design antenna maps that turn AEs on and off in the context of an SM setup. Due to inherent large-array dimensions, a significant number of information bits can be assigned to antenna locations in these maps. However, at very high frequencies, SM becomes challenging because of high channel correlation. 
Adaptive and hierarchical SM solutions can be produced by mapping information bits to antenna indices, at either the SA level or the AE level \cite{Sarieddeen8765243}. The antenna arrays can be visualized as large fully-configurable graphene sheets of AEs that get partially activated. Such arrays can be adapted in real time by activating a specific set of AEs in order to achieve a specific bit rate at a target communication range. Note that SM can be combined with frequency-interleaved antenna-map designs, so as to come up with generic index-modulation solutions. Such solutions can take full advantage of the available resources by assigning information bits to frequency allocations, as well.

\section{Prospective Research Directions}
\label{sec:prop_sol}

Having detailed the channel conditions and research trends, we will discuss a select few prospective THz UM-MIMO use cases that are likely to be realized in the near future. 
In what follows, we promote the use of UM-MIMO-assisted THz communications at the intersection of communications and sensing, as an alternative to wired connectivity in data centers, in the context of large and re-configurable intelligent surfaces, and as an enabler for mobile wireless mid-range communications.

\subsection{Communications and Sensing}
\label{sec:use_nano}
 
Many 6G applications can be piggybacked onto THz wireless communications, particularly in the areas of imaging, localization, and sensing \cite{sarieddeen2019generation}. One interesting application is gas sensing, in which specific spectral absorption characteristics of molecules serve as fingerprints for specific gaseous compositions. UM-MIMO systems can enable sensing over extended distances, where the distance-dependent behavior of molecular absorptions can be mitigated to correct measurements. This mitigation could be exploited to monitor air pollution from a distance. Furthermore, THz signals are used to monitor other physical parameters, such as temperature and displacement. Low-power and low-cost UM-MIMO nano-antenna array configurations can enhance the accuracy of such sensors, by exploiting the spatial degrees of freedom to increase sensing resolution, while at the same time enabling the communication of sensing information over a distributed, wireless-sensing network.

 \subsection{Data Centers}
\label{sec:use_datacenters}

Because of the large number of networked computers and storage devices in data centers, novel communication technologies are required to facilitate the accessing and processing of data. 
Wiring a massive number of servers increases the size of data centers and reduces system efficiency. Wireless links can reduce system costs and yield more energy-efficient data centers by eliminating the need for power-hungry switches. Such links should be complemented by efficient networking solutions and scheduling mechanisms, that allocate channels to servers based on the traffic demand.

High data rates make THz communications a strong candidate for wireless data centers. Furthermore, the reconfigurability of THz antenna arrays can be leveraged to support multiple inter-rack and intra-rack communication links. THz UM-MIMO transceivers with high power, low noise figures, and adequate sensitivity can thus be optimized in such static environments. Moreover, in the particular case of enclosed under-water data centers that use nitrogen for cooling, THz absorption losses are significantly reduced. Note that THz for data centers is already attracting attention; for example, TERAPOD (a Horizon 2020 project supported by the European Union) is using data centers as a proof-of-concept deployment of end-to-end THz wireless links.

 \subsection{Large Intelligent Surfaces}
\label{sec:LINs}

One vision for beyond-5G communication paradigms is to make the entire environment intelligent and active for communication purposes. In this context, the concept of LISs, surfaces that scale up beyond conventional antenna arrays, and act as transmitting and receiving structures in an environment, has been recently proposed. These surfaces should achieve extremely high data rates, support efficient wireless charging capabilities, and enable high-precision sensing applications. In particular, LISs can be realized via THz UM-MIMO because of two main favorable conditions: first, LISs are more likely to yield perfect LoS indoor and outdoor propagation environments; and second, they impose few restrictions on how AEs can be spread. Hence, mutual coupling effects and antenna correlations can be easily avoided. LISs further support simple feedback mechanisms and channel-estimation techniques, which are important for low-latency applications. In addition to active LISs, passive reconfigurable LIS configurations that can modify the signal phase and scale up signal power are also gaining attention. In particular, graphene-based, plasmonic, reconfigurable metasurfaces can achieve beamsteering, beam-focusing, and wave-vorticity control by means of local tuning. Because of the very small size of AEs at THz frequencies, it is difficult to obtain coherent antenna arrays; therefore, the deployment of reconfigurable LISs is a practical solution.

\subsection{Cell-Free Systems}
\label{sec:CF}
\begin{figure}[t]
\centering

\includegraphics[width=3.5in]{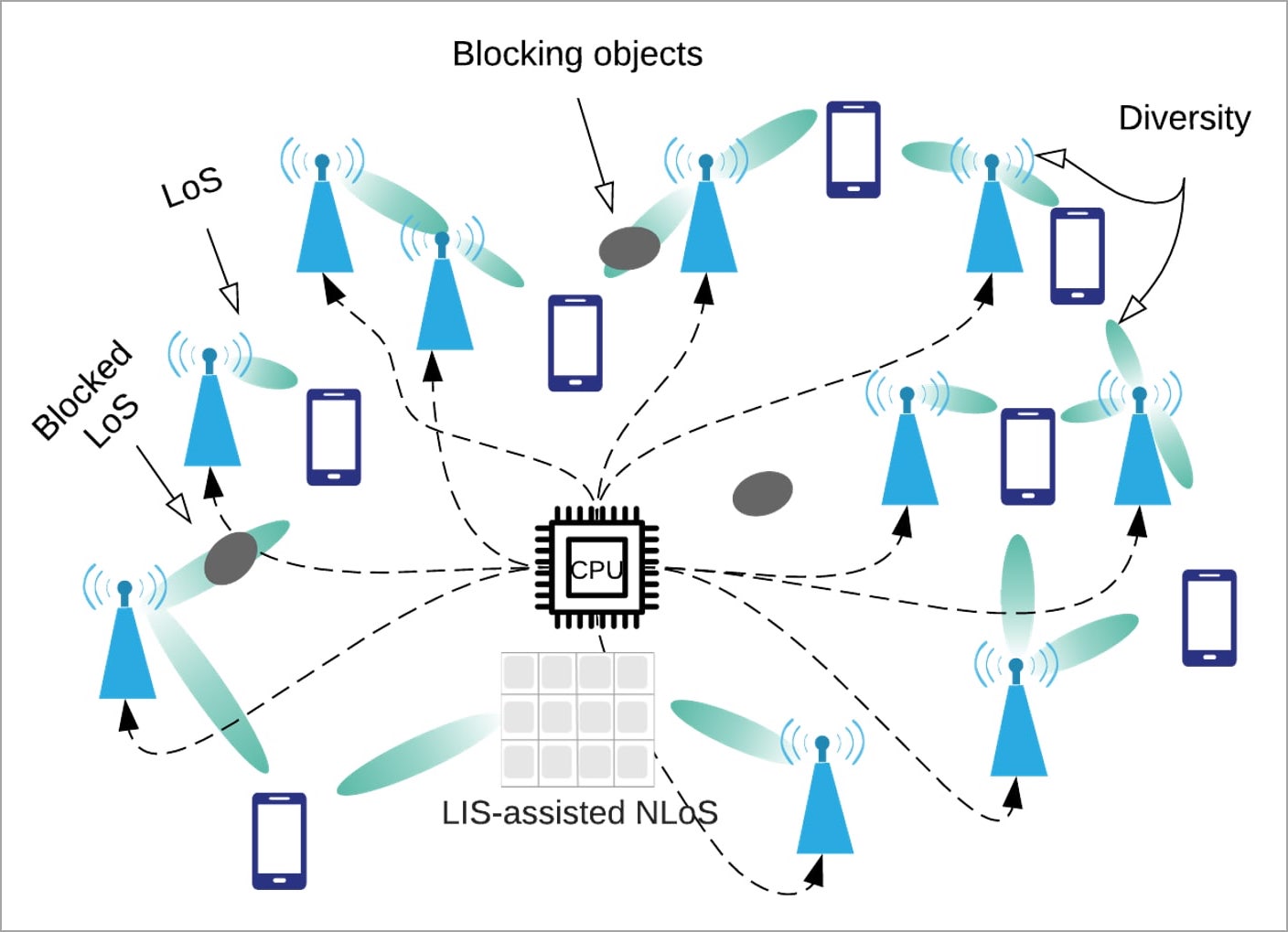}
\caption{An illustration of a cell-free massive MIMO system operating at the THz band.}
\label{f:CF}
\end{figure}

We foresee beyond 5G wireless systems relying heavily on the use of large-scale antenna arrays operating at high frequencies. In conventional massive MIMO systems, a BS with a massive number of antennas simultaneously serves multiple users in a cell. Recently, the concept of cell-free massive MIMO has been proposed to avoid inter-cell interference. While interference is not severe at higher frequencies, cell-free configurations are still important for several other reasons. Cell-free systems comprise a massive number of access points (APs) controlled by a central processing unit. As shown in Fig.\ref{f:CF}, the APs are randomly and densely deployed to serve a much smaller number of users. The THz capabilities can be well-utilized in cell-free systems, in which each user is placed at a short distance from at least one AP so that users can communicate effectively within communication-range limitations. Furthermore, THz-band blockage can easily be mitigated by employing redundant paths in a cell-free network. Conversely, THz communications can in turn enhance the performance of cell-free systems. The deployment of lone distributed massive antenna arrays might fall short of meeting the required data rates and quality-of-service demands; communicating at THz frequencies could thus rectify such shortcomings. Note that the convergence of THz communications, sensing, and localization is favored in a dense, cell-free system. Although cell-free massive MIMO systems at mmWave frequencies have already been studied in the literature \cite{cellfree}, their applicability to the THz band is still an open problem.

\subsection{Mid-Range Mobile Communications}
\label{sec:LANs}

Mid-range wireless mobile-communication applications that require several meters of distance coverage are the holy grail of THz communications and are the main inspiration behind developing UM-MIMO techniques that operate efficiently at the THz band. 6G-and-beyond THz communications promise to support ultra-broadband and ultra-high-speed applications. 
Adaptive and compact THz UM-MIMO array designs allow for sharing transceiver resources, tuning carrier frequencies, and directing antenna beams to multiple users. THz communications bring many exciting opportunities for vehicular networks, as well. In fact, transmitting at high data rates causes systems to be quasi-static, even when users are mobile. Similarly, fast-moving, unmanned aerial vehicles or drones are highly dependent on the throughput, reliability, and latency of wireless systems, making THz UM-MIMO a compelling solution.

One of the main challenges in these mobile setups is to mitigate the effects of blockage. Blockage can easily occur over the medium due to small wavelengths, and at the source due to tiny suspended particles that are big enough to block AEs. Furthermore, since the amount of data that can be transmitted per connection at THz frequencies is potentially huge, the Doppler effect is minimized for an LoS path. The fading-based Doppler spread of multi-path scenarios, however, is much more severe at THz frequencies. While it is not easy to overcome all the challenges that govern THz wireless communications, a plausible solution is to allow for the coexistence of THz, mmWave, and microwave systems. In the mean time, high mobility can still be treated by the more mature mmWave solutions, while backhaul transmissions can be conducted at the THz band.

\section{Conclusion}
\label{sec:conclusion}

In this paper, we examine the characteristics of the THz channel in order to advocate the potential of UM-MIMO systems at high frequencies. With proper configurations, UM-MIMO antenna arrays can overcome distance and power limitations. We argue that graphene-based nano-antenna arrays, in particular, can efficiently realize THz UM-MIMO systems. We define multiple research advances that are critical for increasing the efficiency of THz communications, including signal modulation, waveform design, and resource allocation. Finally, building on all the preceding arguments, we envision a select few 6G-and-beyond use-cases that are likely to realize THz UM-MIMO.

\section{Acknowledgements}
An early version of this work was presented at the 43rd Wireless World Research Forum (WWRF43) meeting in London, UK, Oct. 2019  \cite{faisal2019WWRF}.


\newpage

\section*{Biographies}
\footnotesize

\textbf{Alice Faisal} (S'18) is a senior electrical and computer engineering student at Effat University, Jeddah, Saudi Arabia. She is currently the chair of the IEEE Women in Engineering affinity group at the student branch of Effat University. Her research interests are in the areas of wireless communications and signal processing.

\textbf{Hadi Sarieddeen} (S'13-M'18) received his B.E. degree in computer and communications engineering from Notre Dame University-Louaize, Lebanon, in 2013, and his Ph.D. degree in electrical and computer engineering from the American University of Beirut (AUB), Lebanon, in 2018. He is currently a postdoctoral fellow at King Abdullah University of Science and Technology (KAUST), Thuwal, Saudi Arabia. His research interests are in the areas of wireless communications and signal processing for wireless communications.

\textbf{Hayssam Dahrouj} (S'02, M'11, SM'15) received his computer and communications engineering degree from AUB in 2005, and his Ph.D. degree in electrical and computer engineering from the University of Toronto (UofT) in 2010. In July 2020, he joined the Center of Excellence for NEOM Research at KAUST as a senior research scientist. His main research interests include cloud radio access networks, cross-layer optimization, cooperative networks, convex optimization, distributed algorithms, machine learning, and optical communications networks.

\textbf{Tareq Y. Al-Naffouri} (M'10-SM'18) received his Ph.D. degree in Electrical Engineering from Stanford University in 2004. He is currently a Professor at the Electrical and Computer Engineering department at KAUST. His research interests lie in the areas of sparse, adaptive, and statistical signal processing, localization, machine learning, and their applications.

\textbf{Mohamed-Slim Alouini} (S'94-M'98-SM'03-F'09) was born in Tunis, Tunisia. He received his Ph.D. degree in Electrical Engineering from Caltech, Pasadena, CA, in 1998. He served as a faculty member at the University of Minnesota, Minneapolis, then at Texas A\&M University at Qatar, Education City, Doha, Qatar before joining KAUST as a professor of electrical engineering in 2009. His current research interests include the modeling, design, and performance analysis of wireless communication systems.

\ifCLASSOPTIONcaptionsoff
  \newpage
\fi

\end{document}